\documentclass[twocolumn,superscriptaddress,showpacs,showkeys,preprintnumbers,amsmath,amssymb]{revtex4}

\usepackage{graphicx}
\usepackage{dcolumn}
\usepackage{bm}
\usepackage{longtable}
\begin{document}


\title{Diffusion accompanying noise induced transport in frictional
ratchets}
\author{Raishma Krishnan}
\affiliation{Institute of Physics, Bhubaneswar-751005, India}
\email{raishma@iopb.res.in, jayan@iopb.res.in}
\author{Debasis Dan}
\affiliation{Department of Physics, Indiana University, Bloomington 
47405, U.S.A}
\email{ddan@indiana.edu}
\author {A. M. Jayannavar}
\affiliation{Institute of Physics, Bhubaneswar-751005, India}

\begin{abstract}
We study the noise induced transport of an overdamped 
Brownian particle in frictional ratchet systems in the presence 
of external Gaussian white noise fluctuations. The analytical expressions 
for current and diffusion coefficient are derived and the reliability 
or coherence of transport are discussed by means of their ratio. 
We show that frictional ratchets exhibit larger coherence 
as compared to the  flashing and rocking ratchets.

\end{abstract}
\pacs{05.40.Jc, 05.40.Ca, 02.50.Ey}

\keywords{Ratchets, Brownian motors, Noise, Transport coherence}
\maketitle


\section{Introduction}

The phenomenon of noise induced transport (thermal ratchets) are 
being discussed extensively in recent years. Ratchets are systems that 
exploit the nonequilibrium fluctuations present in the medium 
to generate a directed flow of Brownian particles in 
the absence of a bias~\cite{reiman,1amj}.  Such ratchet models are found to 
have wide range of applications in biological systems and 
in nanotechnology~\cite{appl}.  

Several physical models like 
 flashing ratchets~\cite{ajdari}, rocking ratchets~\cite{magnasco} etc.,
 where potential has been taken to be asymmetric in space, 
 have been developed. In these models to generate noise induced 
 directional transport the nonequilibrium fluctuations need to be 
correlated in time. One can also generate unidirectional currents 
in the presence of symmetric ratchet potentials, however it 
has to be driven by correlated time asymmetric force~\cite{rock-prl}. 
There is yet another class of ratchets, namely the
frictional ratchets which we consider here, where the friction 
coefficient varies in space~\cite{amj,pareek}. 
In such ratchet systems it is possible to 
 get unidirectional currents even in a symmetric potential in 
 the absence of a net bias. Moreover, external fluctuations 
 need not be correlated in time. 
In the presence of external parametric noise 
 the particle on an average absorbs energy from the noise source. 
 The particle spends larger time in the region of space where the 
 friction is higher and hence the energy absorption from the noise 
 source is higher in these regions. Thus the particle in the higher 
 friction regions feel effectively higher temperatures. Hence the problem 
 of particle motion in an inhomogeneous medium in presence of 
 an external noise becomes equivalent to the problem in a space 
 dependent temperature~\cite{amj,pareek,buttiker}. Such systems 
are known to generate unidirectional currents. 
 This  follows as a corollary to  Landauer's blow torch theorem 
that the notion of stability changes 
 dramatically in the presence of temperature inhomogenieties~\cite{land}. 
 In such cases the notion of local stability, valid in equilibrium 
 systems, does not hold. 

Frictional inhomogenieties are common in superlattice structures, 
 semiconductors or motion in porous media. 
Frictional inhomogenity changes the dynamics of the particle nontrivially as 
 compared to the homogeneous case. This in turn has been shown to 
 give rise to many counter intuitive phenomena like 
noise induced stability, stochastic resonance, enhancement in efficiency 
etc.,  in driven non-equilibrium systems~\cite{thesis,luch}.

Extensive studies have been done on the  nature of currents and 
their reversals, stochastic energetics (thermodynamic efficiency) 
etc., in different class of ratchet models. In
contrast, there are very few studies which addresses the question 
of diffusion accompanying transport in ratchet 
systems~\cite{low,sch,ijp}. This is intimately related 
to the question of reliability of transport.
Diffusion infact detriments the quality of transport. In our present 
work we focus on the transport coherence in frictional ratchets 
in the presence of external Gaussian white noise fluctuations. 

Transport of Brownian particles are always associated with   
a diffusive spread. When a particle on an average moves a 
distance $L$ due to its velocity, there will always be an accompanying 
 diffusive spread. If this diffusive spread is much smaller than the 
 distance travelled, then the particle motion is considered to be coherent 
 or optimal or reliable. This is in turn quantified by a 
 dimensionless quantity, Pe$\acute{c}$let 
 number Pe, which is the ratio of current to the diffusion constant. Higher 
 the Pe$\acute{c}$let number, more coherent is the transport. 
Pe$\acute{c}$let number greater than $2$ implies coherence in the
transport~\cite{low}. The Pe$\acute{c}$let numbers for 
some of the models like flashing and rocking ratchets show 
low coherence of transport with $Pe \sim 0.2$ and $Pe \sim 0.6$
~\cite{low} respectively. Another 
study on symmetric periodic potentials along with spatially 
modulated white noise showed a coherent transport with 
Pe$\acute{c}$let number less than $3$. In the same study a 
special kind of strongly asymmetric potential is found to 
increase $Pe$ to $20$ in some range of physical parameters~\cite{sch}. 
Experimental studies on molecular motors showed more reliable transport 
with Pe$\acute{c}$let number ranging from 2 to 6~\cite{high}. 

In our present work we show that system inhomogeneities help in 
enhancing the coherence in the transport depending sensitively on 
the physical parameters.  For this we consider a 
simple spatially symmetric sinusoidal potential. 
Pe of the order of $3$ can be readily obtained. 
The noise strength of the external parametric white noise 
fluctuations play a constructive role in enhancing 
the coherence in transport. As opposed to this, 
temperature (internal fluctuations) degrades the coherence in transport.  

\section{Model:}
 We consider the overdamped dynamics of a Brownian particle moving 
 in a medium with spatially varying frictional coefficient $\eta(q)$ at 
 temperature $T$. Using a  microscopic treatment the 
 Langevin equations for the Brownian 
 particle in a space dependent frictional medium has been obtained 
 earlier \cite{amj,pareek,sancho}. The corresponding overdamped Langevin 
 equation of motion is given by
 \begin{equation}
 {\dot{q}} = {- {\frac{V^\prime(q)}{\eta(q)}}} - {\frac{k_BT \eta^{\prime}(q)}{2
 {[\eta(q)]}^2}} + {\sqrt {\frac {k_BT}{\eta(q)}}}f(t)
 \end{equation}
 with $<f(t)> = 0$, and $ <f(t)f(t^\prime)> = 2 \delta(t-t^\prime)$ 
 where $<...>$ 
 denotes the ensemble average and $q$ the coordinate of the particle.

 The system is then subjected to an external parametric additive 
 white noise fluctuating force $\xi(t)$, so that 
the equation of motion becomes
 \begin{equation}
 \dot{q} = {- \frac{V^\prime(q)}{\eta(q)}} - \frac {k_BT {\eta^\prime}
 (q)}{ 2{[\eta(q)]}^2} + {\sqrt{\frac{k_BT}{\eta(q)}}f(t)} + \xi(t)
 \end{equation}
 with $<\xi(t)>=0$ and $<\xi(t) \xi(t^\prime)> = 2 \Gamma \delta(t-t^\prime)$,
 where $\Gamma$ is the strength of the external white noise $\xi(t)$. 
 The corresponding Fokker-Planck equation is given by \cite{risken}
 \begin{equation}
 \frac {\partial P}{\partial t}= \frac {\partial}{\partial q} 
 \left[{\left\{\frac{V^\prime(q)}{\eta(q)}\right\}P+ \left\{\frac{k_BT}
 {\eta(q)}+\Gamma\right\}\frac{\partial P}{\partial q}}\right]
 \end{equation}
 For periodic functions $V(q)$ and $\eta(q)$ with periodicity $L$, 
 one can readily obtain analytical expression for particle velocity 
 and is given by~\cite{risken}
 \begin{equation}
 v= L\frac{(1-\exp\,[{-2\,\pi\,\delta}])}{\int_0^{2\pi} dy \exp\,[-\psi(y)] 
 \int_{y}^{y+2\pi}dx \frac{\exp\,\,[{\psi(x)}]}{A(x)}}\label{current}
 \end{equation}
 with the generalized potential $\psi(q)$ as
 \begin{equation}
 \psi(q)=\int^{q} dx \frac{V^\prime(x)}{k_BT+\Gamma \eta(x)}\label{eff-pot}
 \end{equation}
 and  $A(q)$ as
 \begin{equation}
 A(q)=\frac{k_BT+\Gamma\eta(q)}{\eta(q)}\label{aq}
 \end{equation}
 with
 \begin{equation}
 \delta = \psi(q)-\psi(q+2\pi) \label{delta}
 \end{equation}
 which inturn determines the effective slope of the generalized 
 potential $\psi(q)$. Hence the sign of $\delta$ gives the 
 direction of current which follows from Eqn~\ref{current}. 

 In our present work we have taken the potential $V(q)=V_0 \sin(q)$
and $\eta(q)=\eta_0[1-\lambda 
\sin(q-\phi)]$, $0\,<\,\lambda\,<\, 1$. The phase lag $\phi$ between 
$V(q)$ and $\eta(q)$ brings in the intrinsic asymmetry in 
the dynamics of the system. The effective potential 
$\psi(q)$, $\delta$ and $A(q)$ are obtained from 
Eqns.~\ref{eff-pot},~\ref{aq}
and \ref{delta}.
 Following references \cite{rec,dan-giant}, one can obtain exact analytical expressions 
 for the diffusion coefficient $D$ and the average velocity $v$ as
 \begin{equation}
 D=\frac{\int_{q_0}^{q_0+L}\frac{ dx}{L}\,A(x)\, {[I_+(x)]}^2 I_-(x)}{\left[{\int_{q_0}
 ^{q_0+L}\frac{dx}{L}I_+(x)}\right]^3}\label{diffusion}
 \end{equation}
 \begin{equation}
 v= \frac{L(1-\exp\,[-L \delta])}
 {{\int_{q_0}^{q_0+L}\,dx \,I_+(x)}} \label{cur}
 \end{equation}
 where $I_+(x)$ and $I_-(x)$ are as given below
 \begin{eqnarray}
 I_+(x)&=& \frac{1}{A(x)}\,\,\exp\,[\psi(x)]\,\int_{x-L}^{x} dy 
 \,\,\exp\,[-\,\psi(y)] \label{iplus} \\
 I_-(x)&=& \exp\,[- \psi(x)] \int_{x}^{x+L} dy\,\, 
 \frac{1}{A(y)}\,\exp\,[\psi(y)]\label{iminus}
 \end{eqnarray}
 $L$ here represents the period of the potential ($=2 \pi$ in our case).
 Now, the time taken for a Brownian particle to travel a distance 
 $L$ is given as $\tau=L/v$ and the spread of the 
particle in the same time is given as $<(\Delta q)^2>=2D\tau$. 
For a reliable transport we require 
 $<(\Delta q)^2>=2D\tau < L^2$. This in turn implies that $Pe=Lv/D >2$ 
 for coherent transport. 

 \section{Results and Discussions}
 The velocity ($v$), diffusion constant ($D$) and the 
 Pe$\acute{c}$let number ($Pe$) are studied  as a function 
of different physical parameters. All the physical quantities are taken 
in dimensionless form. In particular, velocity and diffusion are 
normalized by $(V_0/\eta_0 L)$ and $(V_0/\eta_0)$ 
respectively. Throughout our work we have set $V_0$ and $\eta_0$ 
to be unity. Similarly, $\Gamma$ and $T$ are scaled with
respect to $V_0$ and $V_0/\eta_0$ respectively. 

\begin{figure}[h]
  \centering
  \includegraphics[scale=0.32]{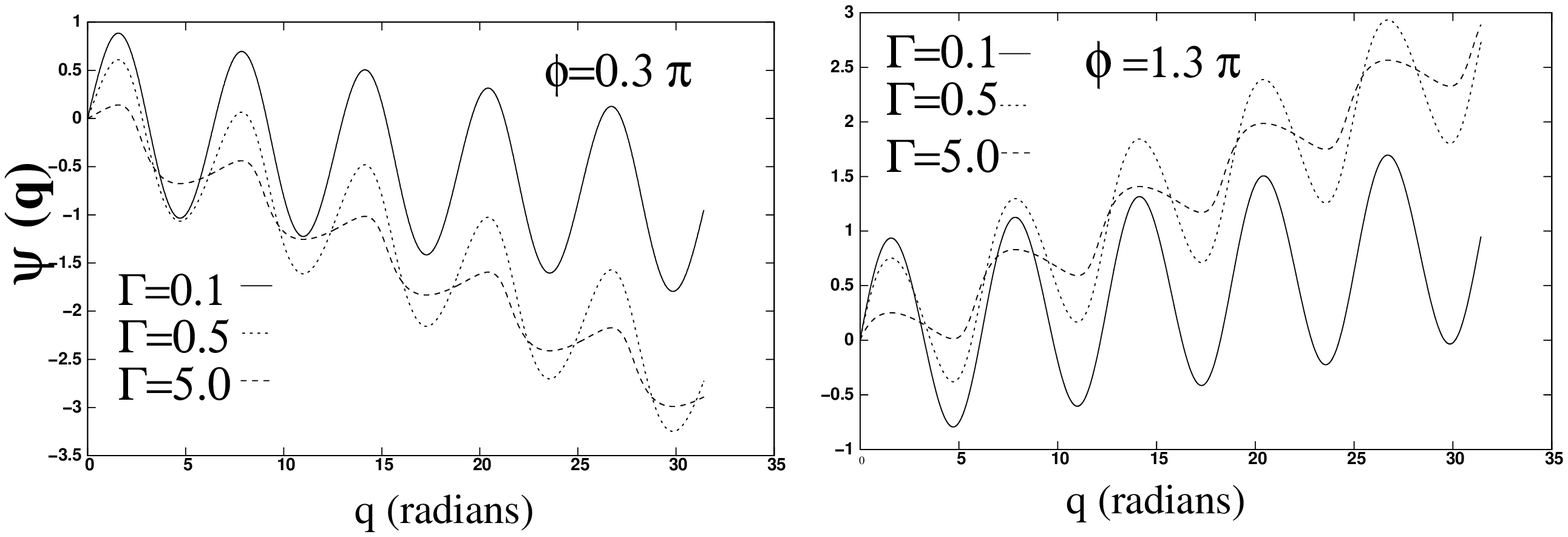}

  \caption{Effective potential $\psi(q)$ as a function of $q$ 
for $V_{0}=1, \lambda=0.9 \mbox{ and } T=1.0$.}
  \label{potential}
\end{figure}
In Fig.~\ref{potential} we have plotted
the effective potential $\psi(q)$ to emphasize this. 
The effective potential is scaled with respect to temperature. 
The nature of currents (in particular the direction of current) 
can be readily inferred from the plot of effective 
potential $\psi(q)$ as a function of $q$ for 
$\phi=0.3\pi$ and $\phi=1.3\pi$ for
various values of physical parameters 
as mentioned in the figure caption. For $\phi$ values 
between $0$ to $\pi$ the effective potential will be tilted 
downward and hence the current direction is positive. 
For $\phi$ values between $\pi$ and $2\pi$ the current is in the 
negative direction. The barrier heights of the effective
potential decreases with increase in the strength 
of the external noise $\Gamma$. This in turn will lead 
to an enhancement in the noise induced velocity which will be 
emphasized in later discussions. 

\begin{figure}[htp!]
 \begin{center}
\input{epsf}
\hskip15cm \epsfxsize=2.7in \epsfbox{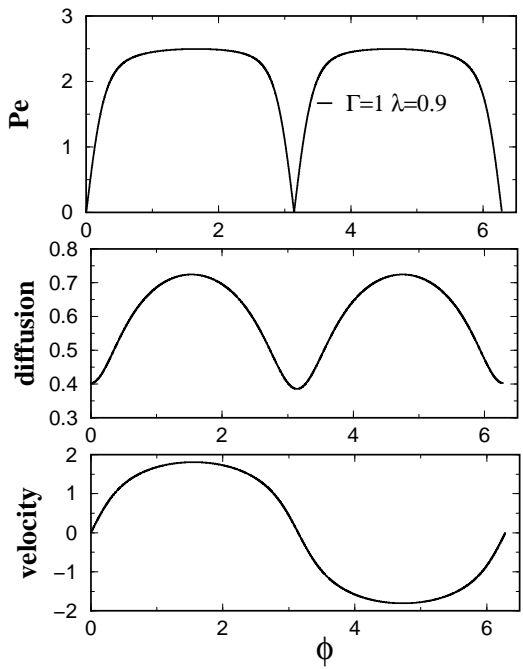}
 \caption{Plot v, D and Pe vs $\phi$ for $\Gamma=1\,,\, 
\lambda=0.9$ and $T=0.01$}. \label{m0vsphi}
 \end{center}
 \end{figure}

In Fig.~\ref{m0vsphi} we plot $v$, $D$ and $Pe$ as a function 
of the phase difference $\phi$ between the periodic functions 
$V(q)$ and $\eta(q)$ for a fixed noise 
 strength, $\Gamma$ and $\lambda$, 
 the amplitude of the periodic modulation of $\eta(q)$. We have 
evaluated these quantities by numerically integrating 
Eqns.~\ref{cur} $\,\&\,$ Eqn.~\ref{iplus} 
 using quadrature methods. All the physical parameters are 
 mentioned in the figure caption. As expected, 
 all the quantities are periodic functions of phase $\phi$ with the 
 velocity $v$ being zero at $\phi=0\,,\pi$ and $2\pi$
~\cite{pareek,buttiker,d1,resonance}.  In 
the range between $0$ to $\pi$ the effective potential $\psi(q)$ will 
be tilted down in the forward direction and the current hence is positive. 
When the phase difference is between $\pi$ and $2 \,\pi$ the effective 
potential will be tilted in the opposite direction and hence 
the current is in the negative direction as discussed 
above. This also follows from the fact that in the region 
of $\phi$ between $0$ to $\pi$ higher friction regions 
lie between the potential minima and the 
corresponding neighbouring maxima on the right and hence the 
particle on the average gains more energy in this 
region as discussed in the introduction. As a consequence the particle 
in a well finds it easier to cross the peak of the potential and go 
over to the right side of the well as compared to crossing over to the other 
side (left of the well). Hence current in the positive direction is 
assured. It should also be noted that when current 
is zero diffusion constant is finite (at the values of 
$\phi=0,\pi$ and $2\pi$)~\cite{amj,buttiker,resonance}.  For the parameters
chosen it appears as if $v$ and $D$ are symmetric around $\pi/2$. 
However it is not so. This has been verified with other parametric values. 
Also, between $0$ and $\pi$ and $\pi$ and $2 \pi$ pe$\acute{c}$let 
number exhibits a maximum with the value being around 
$2.5$. Both the velocity and diffusion
constant exhibit a maxima between $0$ and $\pi$. However, 
the range of variation of the amplitude of velocity is higher than 
that of diffusion resulting in a maxima in pe$\acute{c}$let number.  

\begin{figure}[hbp!]
 \begin{center}
 \input{epsf}
 \hskip15cm\epsfxsize=2.7in \epsfbox{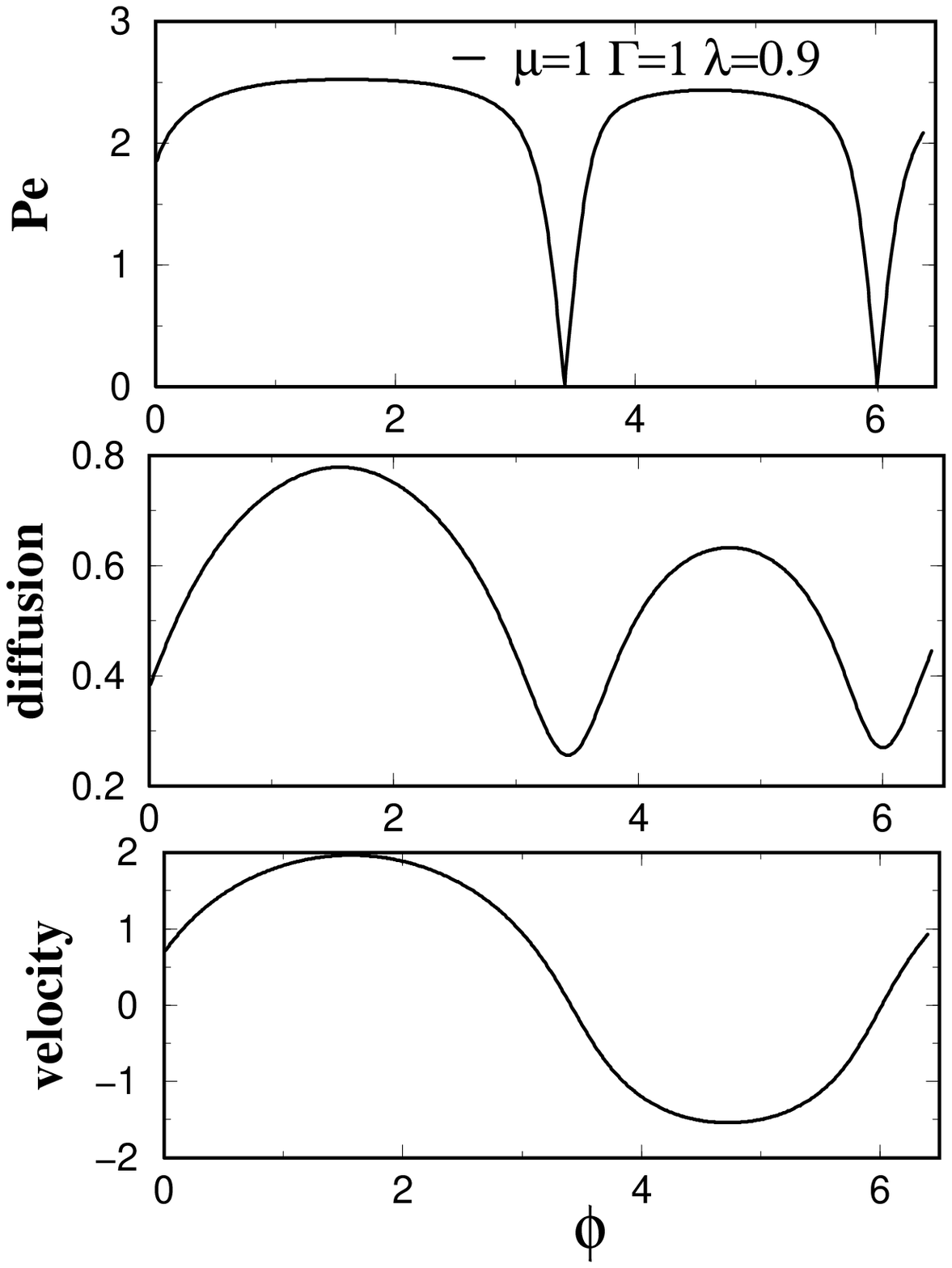}
 \caption{Plot of v, D, and Pe vs $\phi$ for $\mu=1,\,\Gamma=1$, 
$\lambda=0.9$ and $T=0.01$}. \label{m1vsphi} 
\end{center}
\end{figure}
\begin{figure}[hbp!]
 \begin{center}
\input{epsf}
\hskip15cm \epsfxsize=2.7in \epsfbox{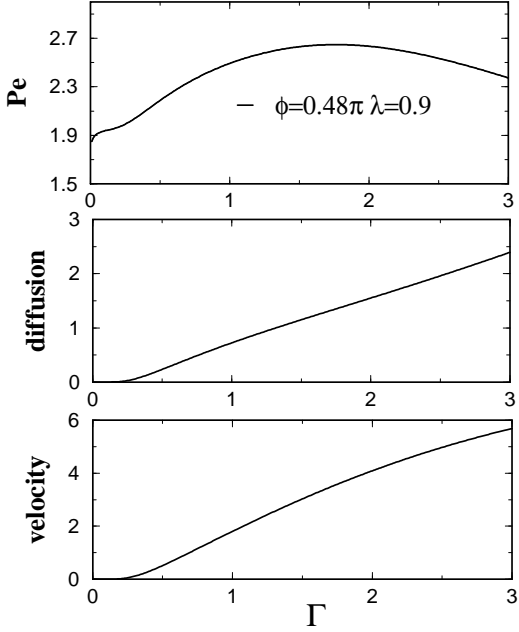}
 \caption{Plot of v, D and Pe vs $\Gamma$ for $\phi=0.48\,\pi\,,\,
\lambda=0.9$ and $T=0.01$}. \label{gamma}
 \end{center}
 \end{figure}

To understand the role of spatial asymmetry in the potential 
we have considered an asymmetric potential of the form  
$V(q)=V_0[\sin(q) - \mu/4 \sin(2q)]$ with the asymmetry 
parameter $\mu=1$. Fig.~\ref{m1vsphi} shows the behaviour 
of $v$, $D$ and $Pe$ for this simple asymmetric case with 
the other physical parameters kept the same as in Fig.~\ref{m0vsphi}. 
For this case we notice that the velocity is not zero when $\phi=0, \pi$ or
$2\pi$ as anticipated. Moreover, it is also clear that the presence of this
simple asymmetry does not make a significant contribution to increase 
an increase in $Pe$. Hence we restrict to the simple 
potential $V=V_0 \sin(x)$ in our further analysis. From these two figures 
we conclude that for a certain range of $\phi$ depending on system parameters
the transport is coherent ($Pe \gg 2$) and moreover this range is quite large.
 In Fig.~\ref{gamma} we plot  $v$, $D$ and $Pe$ as a 
 function of the external white noise strength $\Gamma$. The 
parameters chosen are given in the figure captions. 
The value of $ \phi $ corresponds to the value at which a 
maximum in $Pe$ is seen for the parameters chosen in Fig.~\ref{m0vsphi}.  
We observe that the 
velocity increases monotonically with the external white noise 
and saturates at higher values of $\Gamma $ (not shown in figure). 
In contrast, the diffusion coefficient keeps on increasing indefinitely. The
$Pe$ exhibits a maxima around $\Gamma=1.8$ and the value being 
approximately $2.7$. The reason behind this maxima 
is same as that given in ~\cite{sch} for the
symmetric potential case. 
\begin{figure}[htp!]
 \begin{center}
\input{epsf}
\hskip15cm \epsfxsize=2.7in \epsfbox{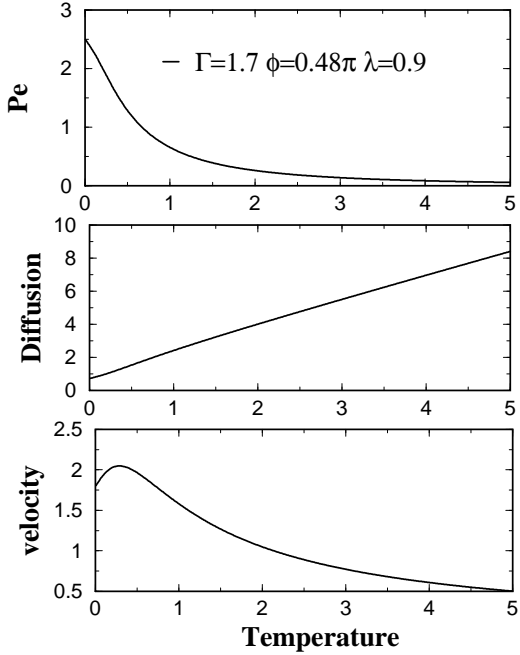}
 \caption{Plot v, D and Pe vs T for $\phi=0.48\,\pi,\,\Gamma=1.7$ and 
$\lambda=0.9$}. \label{temp}
 \end{center}
 \end{figure}
\begin{figure}[htp!]
 \begin{center}
 \input{epsf}
 \hskip15cm \epsfxsize=2.7in \epsfbox{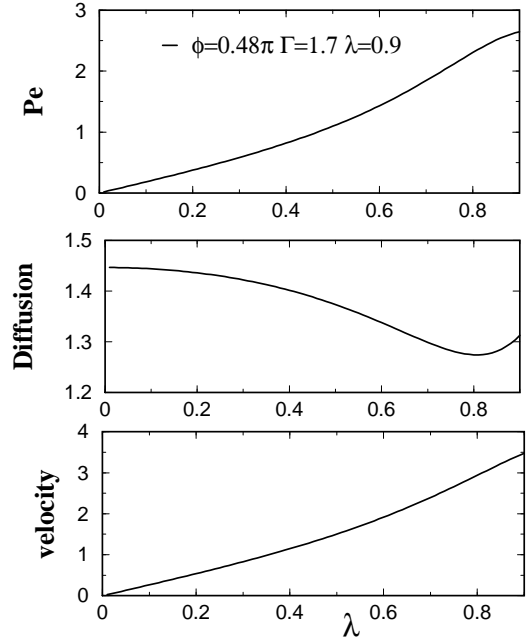}
 \caption{Plot of v, D and Pe vs $\lambda$ for  $\phi=0.48\,\pi,\,
\Gamma=1.7$ and $T=0.01$}. \label{lambda}
 \end{center}
 \end{figure}

In Fig.~\ref{temp} we plot velocity, diffusion and $Pe$ as a function of 
$T$ for $\Gamma = 1.7$.  We observe that 
the current exhibits a peak and monotously decreases to zero as 
a function of $T$. At higher $T$, ($k_BT > \Gamma \eta_0$) 
this behaviour is expected as the temperature overshadows 
the effect of inhomogeneity in the effective potential thus 
suppressing the ratchet effect. If $\Gamma$ is much larger than $1$ 
the peak in the current disappears and hence current monotonously 
decreases with $T$. In contrast, diffusion constant 
increases monotonously. The Pe decreases monotonously with $T$ 
thereby suppressing the coherence in the transport. 

In Fig.~\ref{lambda} we plot velocity, diffusion 
and $Pe$ for values of $\phi = 0.48 \pi$ and $\Gamma=1.7$ 
as a function of the amplitude of oscillation of the 
friction coefficient $\lambda$ ($0<\lambda<1)$. We see that the current
increases as a function of $\lambda$ whereas the diffusion 
constant decreases as a function of $\lambda$. 
In the present case diffusion constant exhibits a
minima near $\lambda=0.8$. However, for other parameter values we see 
that $D$ need not exhibit a minima. It monotonically 
decreases. It should be noted that $\lambda$ cannot be 
greater than or equal to $1$ so as to maintain the friction 
coefficient to be positive. The pe$\acute{c}$let number increases 
monotonously as a function of $\lambda$. This is the only case 
for which we observe a decrease in $D$ while current simultaneously 
increases as  a function of $\lambda$. Thus from our plot we see 
that the presence of $\lambda$ makes the transport more coherent.
 
 \section{Conclusions}

 We have studied the reliability or coherence of transport 
in frictional ratchet systems in the presence of external white noise 
fluctuations. Pe$\acute{c}$let number greater than $2$ can be 
readily obtained in a wide range of parameter space implying 
that the transport is coherent. External noise ($\Gamma$) helps 
in transport coherence whereas the internal noise ($T$) 
degrades the transport coherence. As a function of $\Gamma$ and $T$ 
higher drift velocity is also linked with higher diffusion 
coefficient. It is always desirable to have larger accompanying 
current with minimal diffusive spread to have reliable transport. 
For this we should have regions where velocity increases 
accompanied by a decrease in diffusion coefficient as a 
function of $\Gamma$ and $T$~\cite{sch,dan-giant}. This maybe 
achieved in our model by appropriately choosing strongly 
asymmetric spatially periodic potential like the one 
used in ~\cite{sch}. Our frictional ratchet systems exhibits much 
larger coherence compared to previously studied single particle 
flashing and rocking ratchet systems.


\begin{thebibliography}{0}

 \bibitem{reiman} P. Reimann, Phys. Rep. {\bf 361}, 57 (2002).

 \bibitem{1amj} A. M. Jayannavar, cond-mat 0107079; in Frontiers 
in Condensed Matter Physics, (A commemorative volume to mark 
the $75^{th}$ year of Indian Journal of Physics), ed. J. K. Bhattacharjee 
and B. K. Chakrabarti (in press). 

 \bibitem{appl} Special issue on `Ratchets and Brownian Motors: 
Basic Experiments and Applications', edited by H. Linke, 
Appl. Phys. A 75, 167-352 (2002). 

\bibitem{ajdari} F. J$\ddot{u}$licher, A. Ajdari and J. Prost, 
Rev. Mod. Phys. {\bf 69}, 1269 (1997).

\bibitem{magnasco} M. O. Magnasco, Phys. Rev. Lett. {\bf 72}, 1766 (1994).

\bibitem{rock-prl} A. Ajdari {\it et al.}, J. Phys. I (France) {\bf 4}, 1551 (1994);
M. C. Mahato and A. M. Jayannavar, Phys. Lett. A {\bf 209}, 21 (1995); D. R.
Chialvo and M. M. Millonnas, {\it ibid.} {\bf 209}, 26 (1995).

  \bibitem{amj} A. M. Jayannavar and M. C. Mahato, Pramana-J. Phys. 
 {\bf 45}, 369 (1995).

 \bibitem{pareek} M. C. Mahato, T. P. Pareek and A. M. Jayannavar, 
 Int. J. Mod. Phys. B {\bf 10}, 3857 (1996); M. Millonas, 
Phys. Rev. Lett. {\bf{75}}, 3027 (1995); A. M. Jayannavar, 
Phys. Rev. {\bf E53}, 2957 (1996). 
 
\bibitem{buttiker} M. B$\ddot{u}$ttiker, Z. Phys. B    {\bf 68},161 (1987).

\bibitem {land} R. Landauer, J. Stat. Phys. {\bf 53}, 
 233 (1988). 

\bibitem{thesis} D. Dan, PhD Thesis submitted to 
 Utkal University, Bhubaneswar in 2003; D. Dan and A. M. Jayannavar, 
Int. J. Mod. Phys. B {\bf 14}, 1585 (2000); Phys. Rev. {\bf E60}, 
6421 (1999); D. Dan, M. C. Mahato and A. M. Jayannavar, 
Phys. Rev. {\bf E63}, 056307 (2001); Raishma Krishnan, 
M. C. Mahato and A. M. Jayannavar, Phys. Rev. {\bf E70}, 021102 (2004). 

\bibitem{luch} R. H. Luchsinger, Phys. Rev. {\bf E62}, 272 (2000).  

\bibitem{low} J. A. Freund and L. Schimansky-Geier, Phys. Rev. {\bf E60}, 
 1304 (1999); T. Harms and R. Lipowsky, Phys. Rev. Lett. {\bf{79}}, 
 2895 (1997). 

 \bibitem{sch} B. Linder and L Schimansky-Geier, Phys. Rev. Lett. {\bf 89}, 
 230602 (2002).

\bibitem{ijp} Raishma Krishnan, Debasis Dan and A. M. Jayannavar, 
Ind. J. Phys. {\bf 78}, 747 (2004).

\bibitem{high} M. J. Schnitzer and S. M. Block, Nature (London) 
 {\bf 388}, 386 (1997); K. Visscher, M. J. Schnitzer and S. M. Block, 
 Nature {\bf 400}, 184 (1999).

 \bibitem{sancho} J. M. Sancho, M. San Miguel and D. Duerr, 
J. Stat. Phys. {\bf 28}, 291 (1982).

 \bibitem{risken} H. Risken, 
 The Fokker-Planck Equation (Springer Verlag, Berlin, 1984). 

 
  \bibitem{rec} P. Reimann, C. Van den Broeck, H. Linke, 
 P. H$\ddot{a}$nggi, J. M. Rubi and A. Perez-Madrid, Phys. Rev. 
 {\bf E65}, 31104 (2002); P. Reimann, C. Van den Broeck, H. Linke, 
 P. H$\ddot{a}$nggi, J. M. Rubi and A. Perez-Madrid, 
 Phys. Rev. Lett. {\bf 87},10602 (2001); B. Linder and 
 L. Schimansky-Geier, Fluct. Noise Lett. {\bf 1}, R25 (2001). 
 
\bibitem{dan-giant} D. Dan and A. M. Jayannavar, Phys. Rev. {\bf E66}, 41106 (2002).
 
 
\bibitem{d1}D. Dan, M. C. Mahato and A. M. Jayannavar, Phys. Lett. 
 {\bf 258}, 217 (1999); D. Dan, A. M. Jayannavar and M. C. Mahato, 
 Int. J. Mod. Phys. B {\bf 14} 1585 (2000).
  
\bibitem{resonance}  M. C. Mahato and A. M. Jayannavar, Resonance 
{\bf 8}, No:7, 33 (2003), M. C. Mahato and A. M. Jayannavar, Resonance 
{\bf 8}, No:9, 4 (2003).


 \end{thebibliography}
 \end{document}